\documentclass[aps,reprint,superscriptaddress,showpacs,amssymb,prb,floatfix,preprintnumbers,longbibliography]{revtex4-1}
\usepackage{graphicx,color}
\usepackage{amsmath}
\usepackage{amssymb}
\usepackage{epstopdf}

\newcommand{\slrrtext}{spin-lattice-relaxation rate}
\newcommand{\bacofeasx}{Ba(Fe$_{1-x}$Co$_x$)$_2$As$_2$}
\newcommand{\slrr}{$T_1^{-1}$}

\usepackage[]{hyperref}
\hypersetup{colorlinks=true,linkcolor=blue,citecolor=blue,urlcolor=blue,pdfpagemode=UseNone}
\usepackage{breakurl}

\begin{document}
\thispagestyle{myheadings}


\title{NMR study of nematic spin fluctuations in a detwinned single crystal
of underdoped Ba(Fe$_{1-x}$Co$_{x}$)$_{2}$As$_{2}$}

\author{T. Kissikov}

\affiliation{Department of Physics, University of California, Davis, CA 95616,
USA}

\author{A. P. Dioguardi}

\affiliation{Condensed Matter and Magnet Science, Los Alamos National Laboratory,
Los Alamos, New Mexico 87545, USA}

\author{E. I. Timmons}

\author{M. A. Tanatar}

\author{R. Prozorov}

\author{S. L. Bud'ko}

\author{P. C. Canfield}

\affiliation{Ames Laboratory U.S. DOE and Department of Physics and Astronomy,
Iowa State University, Ames, Iowa 50011, USA}

\author{R. M. Fernandes}

\affiliation{School of Physics and Astronomy, University of Minnesota, Minneapolis,
Minnesota 55116, USA}

\author{N. J. Curro}

\affiliation{Department of Physics, University of California, Davis, CA 95616,
USA}

\date{\today}
\begin{abstract}
We report the experimental details of how mechanical detwinning can
be implemented in tandem with high sensitivity nuclear magnetic resonance
measurements and use this setup to measure the in-plane anisotropy
of the spin-lattice relaxation rate in underdoped Ba(Fe$_{1-x}$Co$_{x}$)$_{2}$As$_{2}$
with $x=0.048$. The anisotropy reaches a maximum of 30\% at $T_{N}$,
and the recovery data reveal that the glassy behavior of the spin
fluctuations present in the twinned state persist in the fully detwinned
crystal. A theoretical model is presented to describe the spin-lattice
relaxation rate in terms of anisotropic nematic spin fluctuations.
\end{abstract}
\maketitle

\section{Introduction}

\label{sec:intro}

The iron pnictide superconductors exhibit rich phase diagrams with
several competing electronic phases.\cite{doping122review,JohnstonPnictideReview}
The parent undoped materials are tetragonal paramagnets at high temperature,
but undergo an orthorhombic structural distortion prior to, or coincident
with, long range antiferromagnetic order of the Fe moments at low
temperature. With electron or hole doping, the structural and magnetic
ordering temperatures are suppressed, and for sufficiently high doping
levels superconductivity emerges, often manifesting a maximum in transition
temperature, $T_{c}$, near the boundary of the orthorhombic/antiferromagnetic
phase. Although this phase diagram is similar to other unconventional
superconductors, a question that has remained open in the pnictides
is whether the antiferromagnetic fluctuations play a role in the superconducting
mechanism, and how these competing phases are related to the orthorhombic
distortion.\cite{uedareview,chubukovreview,KivelsonNematicQCP2015,ScalapinoNematicQCP2015}

The tetragonal-to-orthorhombic transition at $T_{s}$ is driven by
an electronic nematic instability that breaks the $C_{4}$ point group
symmetry of the lattice.\cite{FisherScienceNematic2012,FernandesNematicPnictides,FradkinNematicReview}
The microscopic origin of this nematic instability has been the subject
of intense debate -- in particular, whether it arises from spin or
orbital degrees of freedom (for a review, see Ref. \onlinecite{Fernandes2014}). In the nematic phase, stripe-like
magnetic order sets in at a temperature $T_{N}\leq T_{s}$, with the
moments ordered ferromagnetically along the $b_{{\rm O}}$ direction
and antiferromagnetically along $a_{{\rm O}}>b_{O}$ (see Fig. \ref{fig:xtalphotos}).\cite{Cruz2008,DaiRMP2015} In the absence of strain, orthorhombic
distortions can occur along any of the two degenerate directions,
leading to twin domains in bulk crystals.\cite{prozorovDW122,TanatarTensileStressPRB}
Twinning therefore precludes measurements of anisotropic behavior
in the $ab$ plane, since the crystal contains nominally equal populations
of all such domains. On the other hand, cooling through the structural
transition while maintaining uniaxial stress along the (100)$_{{\rm O}}$
direction can nucleate single domains.\cite{IronArsenideDetwinnedFisherScience2010,TanatarDetwin}
Strain, therefore, provides an avenue to uncover the intrinsic planar
anisotropy in detwinned crystals. Transport and neutron scattering
studies under elastic strain have uncovered large nematic correlations
both in the charge and spin degrees of freedom.\cite{FisherScienceNematic2012,StrainedPnictidesNS2014science,StrainBa122neutronsPRB2015,Lu2016}
These nematic fluctuations diverge at $T_{s}$, and persist well into
the paramagnetic tetragonal phase. Angle-resolved photoemission \cite{ARPESBa122detwinned2011PNAS}
and infrared optical reflection spectroscopy \cite{Mirri2015} studies
have also been conducted on detwinned crystals, revealing a distinct
Fermi surface anisotropy both above and below the structural transition
of the unstrained sample.

Nuclear magnetic resonance (NMR) of the $^{75}$As offers detailed
information about the temperature dependence of these spin fluctuations
and their doping dependence.\cite{takigawaSr122,takigawa2008,imaiBa122overdoped,ImaiLightlyDoped,IyeJPSLorbitalnematicity2015,NakaiPRB2013,NakaiPdopedBa122PRL}
A recent NMR study in underdoped \bacofeasx\ revealed that critical
slowing down sets in at $T_{s}$, however these crystals remained
twinned and therefore no information about the anisotropy of these
critical fluctuations was available.\cite{PhysRevB.89.214511} NMR
studies have also uncovered significant dynamical inhomogeneity in
\bacofeasx, giving rise to stretched exponential relaxation observed
in the \slrrtext.\cite{DioguardiNematicGlass2015,HajoPRB2014,Ba122ClusterGlassNMR,Dioguardi2010}
The crystals in these studies also remained twinned, and an open question
is whether the intrinsic spin fluctuations are sufficiently anisotropic
to explain the broad range of \slrrtext s observed in the presence
of multiple twin domains.

Here we report NMR measurements on a detwinned single crystal of \bacofeasx\ with
$x=0.048$ suspended across a mechanical horseshoe clamp. NMR results
under uniaxial stress in these materials have not been reported previously.
Spin-lattice relaxation rate measurements in twinned samples include
contributions from both domains simultaneously, and therefore the
magnetization recovery may consist of a distribution of relaxation
rates.\cite{Ba122ClusterGlassNMR} The material studied here is underdoped,
with $T_{s}\approx60$ K, $T_{N}\approx50$ K, and $T_{c}\approx18$
K. We find that the spin lattice relaxation rate is anisotropic in
the basal plane, reflecting strong nematic spin correlations of the
Fe spins extending above $T_{s}$. We also find that the stretched
exponential recovery persists in the detwinned crystals. These results
suggest that random strain fields induced by the dopants is greater
than the externally applied strain used to detwin the crystal. The
paper is organized as follows: Section \ref{sec:strain_and_resistivity} describes the strain device and resistivity measurements, Section \ref{sec:T1} describes the spin-lattice relaxation measurements, and Section \ref{sec:analysis} describes the interpretation of the relaxation rate data in terms of nematic spin fluctuations. Details of the calculation of \slrr\ in terms of the dynamical spin susceptibility are given in the appendix.


\section{Application of strain and detwinning}

\label{sec:strain_and_resistivity}

\begin{figure}
\includegraphics[width=1\linewidth]{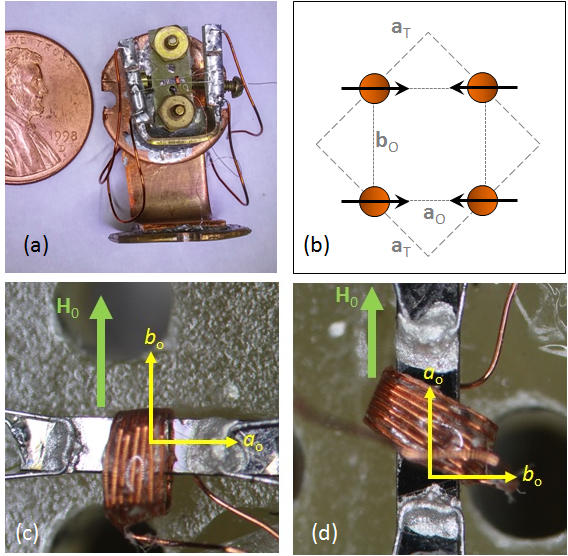} \protect\protect\caption{(a) Strain device with single crystal of Ba(Fe$_{0.952}$Co$_{0.048}$)$_{2}$As$_{2}$
under strain. (b) $ab$ plane, showing the Fe atoms and spin orientation
in the ordered magnetic phase, with the orthorhombic $a_{{\rm O}}$
and $b_{{\rm O}}$ axes shown as dotted lines, and the tetragonal
axes ($a_{T}$) shown as dashed lines. (c) Close-up image of the crystal
oriented such that the applied field, $\mathbf{H}_{0}$, is along
the $b_{{\rm O}}$ (perpendicular to the direction of applied strain)
and (d) along $a_{{\rm O}}$ (parallel to the direction of applied
strain). For the latter case, the coil was rotated by approximately
30$^{\circ}$ so that a component of the radiofrequency field $\mathbf{H}_{1}$
lies perpendicular to $\mathbf{H}_{0}$.}

\label{fig:xtalphotos}
\end{figure}

Single crystals were synthesized via a self-flux method and characterized
via transport measurements and wavelength-dispersive spectroscopy
to determine the Co-doping level.\cite{CanfieldBa122phasediagram2008}
A sample of dimensions 1.1mm $\times$ 0.57 mm $\times$ 0.05 mm was cut with the long axis
parallel to the tetragonal {[}110{]} direction, and mounted in a mechanical
horseshoe device as described in Ref. \onlinecite{TanatarDetwin} and shown in
Fig. \ref{fig:xtalphotos}. The crystal was secured using silver wires
soldered to the edges of the sample. These wires serve not only to
transmit tensile stress to the crystal but also as leads for resistivity
measurements. Stress is applied by tightening a screw by about 1/4
to 1/2 turn, which is enough to apply stresses on the order of 10-20
MPa.\cite{TanatarDetwin} The sample was inserted into the NMR coil
embedded in epoxy prior to mounting in the clamp cell. This is the
first time such a device has been employed for NMR measurements.

\begin{figure}
\includegraphics[width=1\linewidth]{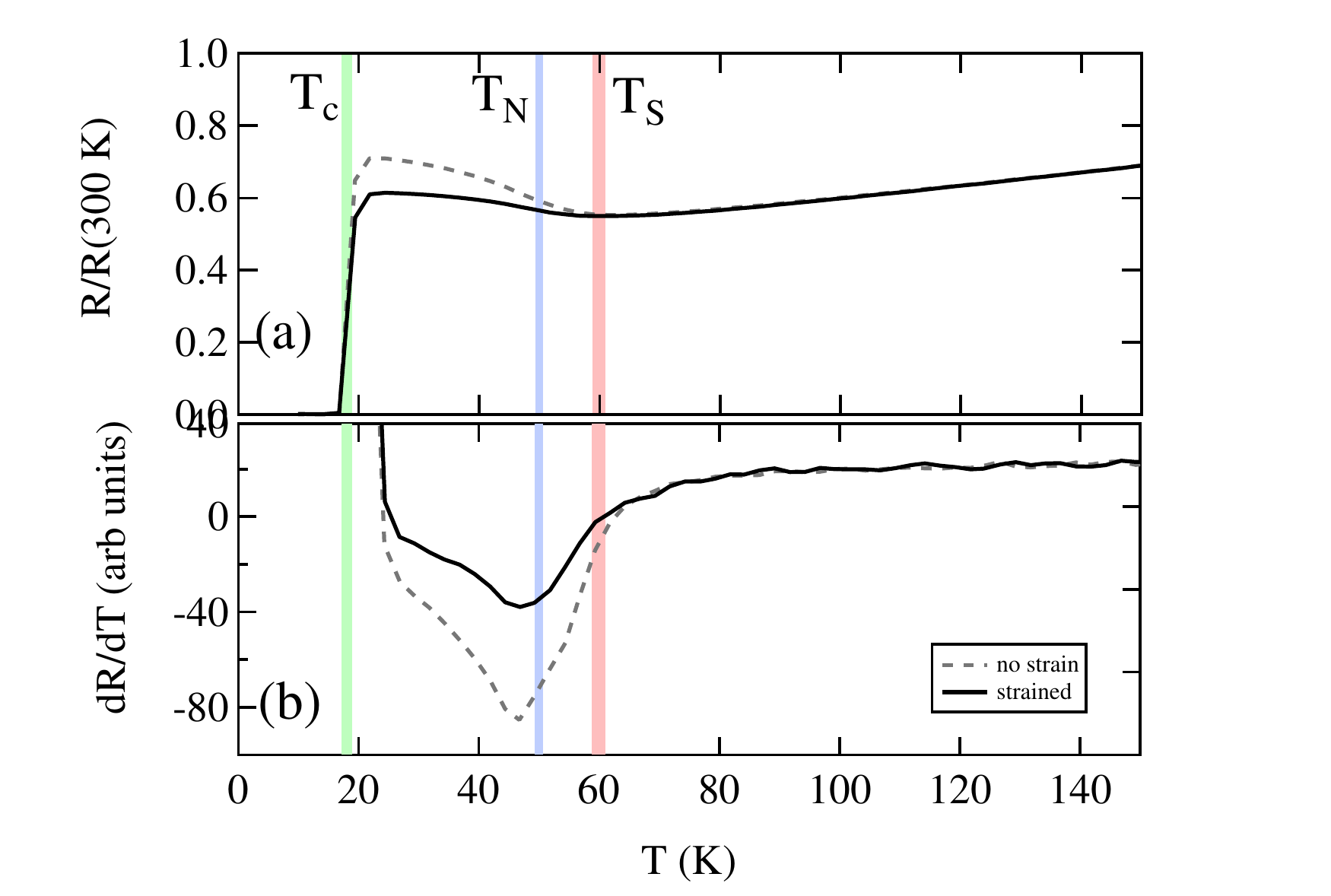} \protect\protect\caption{(a) Resistance and (b) derivative of resistance versus temperature
for the Ba(Fe$_{0.952}$Co$_{0.048}$)$_{2}$As$_{2}$ crystal measured
with and without strain in zero field. 
resistivity reaches a minimum at $T_{S}$, and the $dR/dT$ curve
exhibits a minimum at $T_{N}$.\cite{doping122review} }
\label{fig:resistivity}
\end{figure}

The resistance of the crystal is shown in Fig. \ref{fig:resistivity}
as a function of temperature measured in zero magnetic field. In the
unstrained state, the resistivity exhibits a minimum at $T_{s}$,
and the temperature derivative $dR/dT$ curve exhibits a broad maximum
close to $T_{N}$.\cite{CanfieldBa122phasediagram2008,doping122review}
Note, however, that we identify $T_{N}$ not by the resistance measurements,
but by the peak in \slrr, as discussed below, since that indicates
a divergence in the critical spin fluctuations. In the absence of
strain, the resistance includes domains oriented both along the crystallographic
$a_{{\rm O}}$ and $b_{{\rm O}}$ directions. Under strain, domains
oriented with the $a_{{\rm O}}$ axis parallel to the direction of
applied tensile strain are favored, and the measured resistance changes
below $T_{s}$. For sufficiently large strain, the measured resistance
becomes independent of strain, indicating a fully detwinned state.
Fig. \ref{fig:resistivity} shows the resistance for the fully detwinned
state. This behavior is consistent with independent measurements of
the resistivity along the $a_{{\rm O}}$ direction.\cite{IronArsenideDetwinnedFisherScience2010}

\section{Spin-lattice relaxation}

\label{sec:T1}

\begin{figure}
\includegraphics[width=1\linewidth]{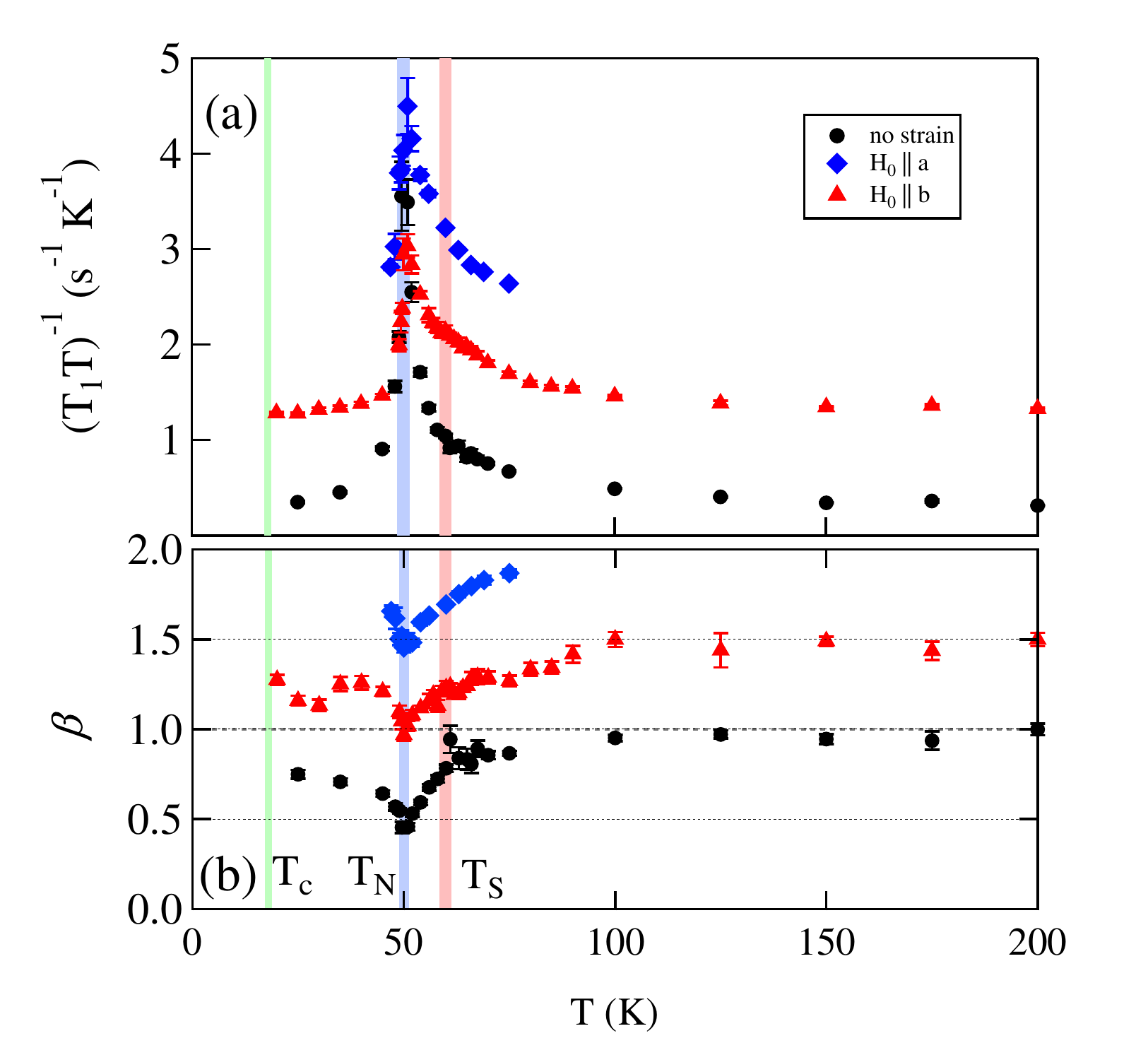} \protect\protect\caption{(a) $(T_1T)^{-1}$ and (b) the stretching exponent $\beta$ versus temperature
for Ba(Fe$_{0.952}$Co$_{0.048}$)$_{2}$As$_{2}$ for field oriented
along either the $a_{{\rm O}}$ or $b_{{\rm O}}$ directions. The
data in (a) have been offset vertically by 1 and 2 s$^{-1}$K$^{-1}$ for
clarity, and in (b) by 0.5 and 1.0.}
\label{fig:rawT1data}
\end{figure}

The \slrrtext, \slrr, was measured at the central transition of
the $^{75}$As ($I=3/2$) by inversion recovery in a field of $H_{0}=11.73$
T for the field perpendicular to the $c$-axis, with and without strain
applied. The measurements were conducted with $\mathbf{H}_{0}$ oriented
both parallel ($\mathbf{H}_{0}~||~\hat{a}_{0}$) and perpendicular
($\mathbf{H}_{0}~||~\hat{b}_{0}$) to the direction of applied strain.
The nuclear magnetization was fit to a stretched exponential, $M(t)=M_{0}\left[1-f(9\exp(-(6t/T_{1})^{\beta})/10+\exp(-(t/T_{1})^{\beta}))/10\right]$,
where $M_{0}$ is the equilibrium magnetization, $f$ is the inversion
fraction, and $\beta$ is the stretching exponent.\cite{Ba122ClusterGlassNMR} $(T_1T)^{-1}$ and $\beta$ are shown as a
function of temperature in Fig. \ref{fig:rawT1data} (note that the
data have been offset vertically for clarity). For  $\mathbf{H}_{0}~||~\hat{b}_{0}$,
the coil was naturally oriented such that the rf field, $\mathbf{H}_{1}\perp\mathbf{H}_{0}$,
as shown in Fig. \ref{fig:xtalphotos}(c); this condition is necessary
in order to detect the nuclear magnetization. For $\mathbf{H}_{0}~||~\hat{a}_{0}$
the coil was rotated by $\sim30^{\circ}$ from the strain
axis as shown in Fig. \ref{fig:xtalphotos}(d) in order to create
a component of $\mathbf{H}_{1}$ that is perpendicular to $\mathbf{H}_{0}$.
The clamp and suspended crystal were warmed to room temperature and
rotated between Figs. \ref{fig:xtalphotos}c and \ref{fig:xtalphotos}d.
Because the applied stress was not changed, the level of
strain was nominally identical for the two orientations. The component
parallel to $\mathbf{H}_{0}$ has no effect on the nuclear magnetization
and does not affect the \slrr\ measurement.

As seen in Fig. \ref{fig:rawT1data}, the relaxation rate diverges
at $T_{N}$, and the stretching exponent, $\beta$, reaches a minimum
of $\approx0.5$ at this temperature. The same qualitative behavior
is observed with and without strain, but there are subtle differences
in $(T_1T)^{-1}$ that emerge near $T_{s}$ under strain. The peak value
of $(T_1T)^{-1}$ decreases by $\sim30$\% for both directions under strain.
Furthermore, the data for $\mathbf{H}_{0}~||~\hat{b}_{0}$ appears to exhibit a
small shoulder at $T_{S}$, that does not appear in the data for
the $a_{{\rm O}}$ direction. Surprisingly $\beta$ does not show
any significant differences under strain. $\beta$ is a direct measure
of the width of the distribution of local relaxation rates.\cite{johnstonstretched}
This distribution has been postulated to arise from random strain fields induced
by the dopants that couple to nematic order, causing $\beta$ to decreases from unity below a temperature
on the order of 100 K in this doping range.\cite{DioguardiNematicGlass2015}
This inhomogeneity might be expected to vanish in the presence of
a homogeneous strain field that is enough to induce a single nematic
domain. However the data indicate that the level of inhomogeneity,
as measured by the size of $\beta$, remains unchanged. This result
suggests that either the origin of the inhomogeneous relaxation arises
from some other source of disorder, or that the random strain fields
induced by the Co dopant atoms,\cite{LoucaPRB2011} which are much larger than the modest
homogenous strain field that is applied to detwin the crystal,
are responsible for the glassy behavior.

Fig. \ref{fig:T1anisotropy}(a) shows the difference $\Delta(T_{1}T)_{\alpha}^{-1}=(T_{1}T)_{\alpha}^{-1}(\varepsilon)-(T_{1}T)_{\alpha}^{-1}(0)$
($\alpha = a,b)$ between the relaxation rates with and without uniaxial tensile strain
for both directions. Fig. \ref{fig:T1anisotropy}(b) shows the anisotropy
in the relaxation rate, $(T_{1}T)_{{\rm anis}}^{-1}=(T_{1}T)_{a}^{-1}-(T_{1}T)_{b}^{-1}$
under strain, and the isotropic strain-induced component, $(T_{1}T)_{{\rm iso}}^{-1}=\frac{1}{2}\left((T_{1}T)_{a}^{-1}(\epsilon)+(T_{1}T)_{b}^{-1}(\epsilon)\right)-(T_{1}T)^{-1}(0)$.
The relaxation was measured for both crystal directions in the absence
of strain, and no differences were observed to within the error bars.
All of these quantities peak at $T_{N}$, but remain finite up to
and above $T_{s}$. This behavior reflects the fact that $C_{4}$
symmetry is broken by the strain field, which induces a finite nematicity
above the onset of long-range nematic order, similarly to how a magnetic
field induces a finite magnetization in the paramagnetic phase above
the Curie temperature in a ferromagnet. Similar behavior has been
observed in elastoresistance and neutron scattering measurements.\cite{FisherScienceNematic2012,StrainBa122neutronsPRB2015,StrainedPnictidesNS2014science}
Note that the magnitude of $(T_{1}T)_{{\rm anis}}^{-1}$ in the detwinned
state is approximately 30\% of the value of $(T_{1}T)^{-1}$ in the
unstrained state at $T_{N}$. The width of the distribution of relaxation rates, however, far exceeds
this variation due to the anisotropy, which is consistent with the
observation that $\beta$ is unchanged in the detwinned state.

\section{Analysis}
\label{sec:analysis}

\begin{figure}
\includegraphics[width=1\linewidth]{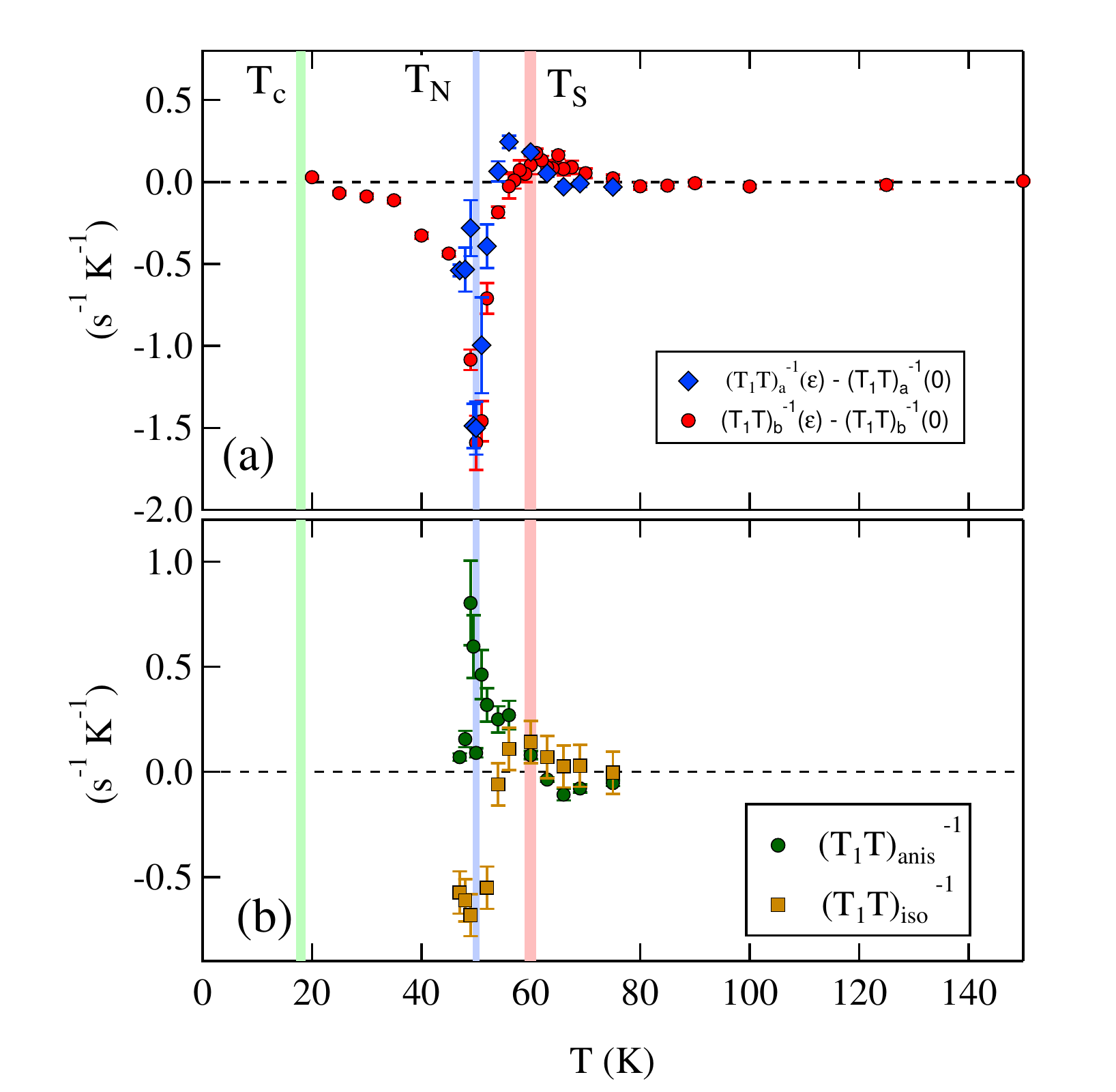} \protect\protect\caption{(a) $(T_{1}T)^{-1}(\epsilon)-(T_{1}T)^{-1}(0)$ for field along both
the $a$ and $b$ directions, and (b) $(T_{1}T)_{{\rm anis}}^{-1}$
and $(T_{1}T)_{{\rm iso}}^{-1}$ versus temperature for Ba(Fe$_{0.952}$Co$_{0.048}$)$_{2}$As$_{2}$.}

\label{fig:T1anisotropy}
\end{figure}

To analyze the results, we start with the general expression for the
spin-lattice relaxation rate due to a magnetic field applied in an
arbitrary direction. Hereafter, our coordinate system refers to the
1-Fe unit cell in the orthorhombic phase. In the paramagnetic state,
the internal field experienced by the nucleus is zero and we obtain
\cite{T1formfactorsArsenides}:

\begin{equation}
\left(\frac{1}{T_{1}T}\right)_{\theta,\phi}=\frac{g^{2}}{2}\sum_{\mathbf{q}}\sum_{i=1,2}\left[\bar{R}\cdot\bar{A}_{\mathbf{q}}\cdot\bar{\tilde{\chi}}\left(\mathbf{q}\right)\cdot\bar{A}_{\mathbf{q}}^{\dagger}\cdot\bar{R}^{\dagger}\right]_{ii}\label{def}
\end{equation}
where $g$ is a constant proportional to the gyromagnetic ratio of
the nucleus, $\bar{A}_{\mathbf{q}}$ is the hyperfine tensor (shown
explicitly in the Appendix), $\bar{R}$ is the rotation matrix (shown
explicitly in the Appendix), and $\theta,\varphi$ are the polar and
azimuthal angles describing the direction of the magnetic field with
respect to the $(a,b,c)$ crystal axis. In this coordinate system,
the susceptibility is diagonal in spin space, $\bar{\chi}=\mathrm{diag}\left(\chi_{aa},\chi_{bb},\chi_{cc}\right)$.
For convenience, we defined $\tilde{\chi}_{\alpha\beta}\left(\mathbf{q}\right)=\lim\limits _{\omega\rightarrow0}\frac{\mathrm{Im}\chi_{\alpha\beta}\left(\mathbf{q},\omega\right)}{\omega}$.
Because the system is metallic, Landau damping $\gamma$ is present
and we can write $\chi_{\alpha\beta}^{-1}\left(\mathbf{q},\omega\right)=\chi_{\alpha\beta}^{-1}\left(\mathbf{q}\right)-i\gamma\omega$,
yielding $\tilde{\chi}_{\alpha\beta}\left(\mathbf{q}\right)=\gamma\chi_{\alpha\beta}^{2}\left(\mathbf{q}\right)$.

The anisotropy in the spin-lattice relaxation rate $\left(\frac{1}{T_{1}T}\right)_{\mathrm{anis}}\equiv\left(\frac{1}{T_{1}T}\right)_{a}-\left(\frac{1}{T_{1}T}\right)_{b}\equiv\left(\frac{1}{T_{1}T}\right)_{\phi=0,\theta=\pi/2}-\left(\frac{1}{T_{1}T}\right)_{\phi=\pi/2,\theta=\pi/2}$
can be calculated directly from Eq. (\ref{def}). In general, the
anisotropy in $1/T_{1}T$ can arise from two sources: either an anisotropy
in the elements of the hyperfine tensor $\bar{A}_{\mathbf{q}}$ or
an anisotropy in the elements of the susceptibility tensor $\chi_{\alpha\beta}$.
The latter reflects the anisotropies of the magnetic fluctuations,
whereas the former is mainly determined by the changes in the lattice
environment. Since the lattice distortions are very small, hereafter
we focus on the anisotropies induced by the spin spectrum only, setting
$A_{aa}=A_{bb}$, $A_{bc}=A_{ac}$, and $A_{ab}=A_{ba}$. We obtain

\begin{align}
\left(\frac{1}{T_{1}T}\right)_{\mathrm{anis}} & =\frac{g^{2}}{2}\sum_{\mathbf{q}}\left\{ \left[F_{1}\left(\mathbf{q}\right)A_{ab}^{2}-F_{2}\left(\mathbf{q}\right)A_{aa}^{2}\right]\times\right.\label{T1T_aniso}\\
 & \left.\left[\tilde{\chi}_{aa}\left(\mathbf{q}\right)-\tilde{\chi}_{bb}\left(\mathbf{q}\right)\right]+F_{3}\left(\mathbf{q}\right)A_{ac}^{2}\tilde{\chi}_{cc}\left(\mathbf{q}\right)\right\} \nonumber
\end{align}
with the form factors $F_{1}\left(\mathbf{q}\right)=\sin^{2}\left(\frac{q_{x}}{2}\right)\sin^{2}\left(\frac{q_{y}}{2}\right)$,
$F_{2}\left(\mathbf{q}\right)=\cos^{2}\left(\frac{q_{x}}{2}\right)\cos^{2}\left(\frac{q_{y}}{2}\right)$,
and $F_{3}\left(\mathbf{q}\right)=\frac{1}{2}\left(\cos q_{x}-\cos q_{y}\right)$.
The existence of a sizable spin-orbit coupling in the iron pnictides
\cite{Borisenko2015} enforces important symmetry constraints on the
susceptibility tensor.\cite{Vafek13,Christensen15} Specifically,
in the tetragonal paramagnetic phase, while $\chi_{aa}\left(\mathbf{q}\right)$
and $\chi_{bb}\left(\mathbf{q}\right)$ do not need to be $C_{4}$
(tetragonal) symmetric functions, $\chi_{aa}\left(\mathbf{q}\right)$
becomes identical to $\chi_{bb}\left(\mathbf{q}\right)$ upon a $90^{\circ}$
rotation. Therefore, because the combination $\tilde{\chi}_{aa}\left(\mathbf{q}\right)-\tilde{\chi}_{bb}\left(\mathbf{q}\right)$
is $C_{2}$ symmetric, while the functions $F_{1}\left(\mathbf{q}\right)$
and $F_{2}\left(\mathbf{q}\right)$ are $C_{4}$ symmetric, the first
term in Eq. (\ref{T1T_aniso}) vanishes in the tetragonal phase. Similarly,
symmetry requires that $\chi_{cc}\left(\mathbf{q}\right)$ is a $C_{4}$
symmetric function; thus, because $F_{3}\left(\mathbf{q}\right)$
is $C_{2}$ symmetric, the second term vanishes as well. Hence, as
expected, $\left(T_{1}T\right)_{\mathrm{anis}}^{-1}$ vanishes in
the tetragonal phase.

To model the magnetic spectrum of the pnictides, we note that at low
energies the susceptibility is strongly peaked at the magnetic ordering
vectors $\mathbf{Q}_{1}=\left(\pi,0\right)$ and $\mathbf{Q}_{2}=\left(0,\pi\right)$,
as seen by neutron scattering.\cite{Inosov2010,Tucker2012}  If
the susceptibilities were delta functions peaked at the ordering vectors,
then the fact that $F_{1}\left(\mathbf{Q}_{i}\right)=F_{2}\left(\mathbf{Q}_{i}\right)=0\neq F_{3}\left(\mathbf{Q}_{i}\right)$
would imply that $\left(T_{1}T\right)_{\mathrm{anis}}^{-1}$ probes
only the $\chi_{cc}$ component of the susceptibility tensor. However,
the system has a finite magnetic correlation length in the paramagnetic
phase. To model this behavior, we consider a low-energy model in which
the susceptibilities are peaked at the magnetic ordering vectors\cite{Fernandes2012}:

\begin{align}
\frac{\chi_{aa}\left(\mathbf{q}\right)}{\chi_{0}} & =\left(\xi_{1}^{-2}+\cos q_{x}-\cos q_{y}+2\right)^{-1}+\nonumber \\
 & \left(\xi_{2}^{-2}-\cos q_{x}+\cos q_{y}+2\right)^{-1}\nonumber \\
\frac{\chi_{bb}\left(\mathbf{q}\right)}{\chi_{0}} & =\left(\xi_{2}^{-2}+\cos q_{x}-\cos q_{y}+2\right)^{-1}+\nonumber \\
 & \left(\xi_{1}^{-2}-\cos q_{x}+\cos q_{y}+2\right)^{-1}\nonumber \\
\frac{\chi_{cc}\left(\mathbf{q}\right)}{\chi_{0}} & =\left(\xi_{3}^{-2}+\cos q_{x}-\cos q_{y}+2\right)^{-1}+\nonumber \\
 & \left(\xi_{3}^{-2}-\cos q_{x}+\cos q_{y}+2\right)^{-1}\label{suscept}
\end{align}
where we defined the correlation lengths $\xi_{i}$ associated with
each magnetic channel (in units of the lattice constant $a$) and
the magnetic energy scale $\chi_{0}^{-1}$. In the tetragonal phase,
symmetry requires that $\chi_{aa}\left(\mathbf{Q}_{1}\right)=\chi_{bb}\left(\mathbf{Q}_{2}\right)$,
$\chi_{bb}\left(\mathbf{Q}_{1}\right)=\chi_{aa}\left(\mathbf{Q}_{2}\right)$,
and $\chi_{cc}\left(\mathbf{Q}_{1}\right)=\chi_{cc}\left(\mathbf{Q}_{2}\right)$
\cite{Christensen15}, which is satisfied by Eq. (\ref{suscept}).
The situation is different in the nematic phase, where magnetic fluctuations
become anisotropic, i.e. fluctuations around $\mathbf{Q}_{1}$ and
$\mathbf{Q}_{2}$ are no longer equivalent. Because the susceptibility
tensor has three independent elements, one needs to introduce three
``nematic order parameters'' $\varphi_{i}$, with $i=1,2,3$. We
introduce them in Eq. (\ref{suscept}) by replacing the magnetic correlation
lengths $\xi_{1}^{-2}\rightarrow\xi_{1}^{-2}\mp\varphi_{1}$, $\xi_{2}^{-2}\rightarrow\xi_{2}^{-2}\pm\varphi_{2}$
(where the upper sign refers to $\chi_{aa}$ whereas the lower sign
refers to $\chi_{bb}$ ), and $\xi_{3}^{-2}\rightarrow\xi_{3}^{-2}\mp\varphi_{3}$
(where the upper sign corresponds to the first term in $\chi_{cc}$
whereas the lower sign refers to the second term). The physical meaning
of these nematic order parameters is clear,\cite{FernandesNematicPnictides}
as $\varphi_{i}>0$ $(\varphi_{i}<0)$ implies that the $\mathbf{Q}_{1}$
$(\mathbf{Q}_{2})$ ordering vector is selected in the nematic phase.
The fact that these three order parameters break the same symmetry
implies that they are either all zero or all non-zero (i.e. $\varphi_{1}\propto\varphi_{2}\propto\varphi_{3}$),
however their relative signs depend on microscopic considerations.

Substituting these expressions in Eq. (\ref{T1T_aniso}) and expanding
to leading order in the three nematic order parameters, we obtain
(in units of $\frac{g^{2}\gamma\chi_{0}}{\pi}$):
\begin{equation}
(T_{1}T)_{{\rm anis}}^{-1}=-2(A_{aa}^{2}-A_{ab}^{2})(\varphi_{1}\xi_{1}^{2}-\varphi_{2}\xi_{2}^{2})-8A_{ac}^{2}\varphi_{3}\xi_{3}^{4}\label{eqn:T1TinvANIS}
\end{equation}
and
\begin{eqnarray}
\Delta(T_{1}T)_{{\rm iso}}^{-1} & = & 8A_{ac}^{2}\xi_{1}^{6}\varphi_{1}^{2}+\frac{1}{2}\left(A_{aa}^{2}+A_{ab}^{2}\right)\xi_{2}^{4}\varphi_{2}^{2}\nonumber \\
 &  & +4A_{ac}^{2}\xi_{3}^{6}\varphi_{3}^{2}.
 \label{eqn:T1TinvISO}
\end{eqnarray}

%

In deriving these expressions, we considered $\xi_{i}$ to be moderately
large and kept the leading order terms for each $\xi_{i}$. We also
neglected any strain-induced changes to the tetragonal hyperfine coupling
tensor. As expected by symmetry consideration, $(T_{1}T)_{{\rm anis}}^{-1}$
varies linearly with $\varphi_{i}$, whereas $\Delta(T_{1}T)_{{\rm iso}}^{-1}$
varies quadratically. According to previous NMR investigations, $A_{aa}\approx0.66$
kOe/$\mu_{B}$ and $A_{ac}\approx0.43$ kOe/$\mu_{B}$ \cite{takigawa2008}.
We do not have direct information about $A_{ab}$, however all of the other elements of the hyperfine tensor are known.  If we assume that one of the principal axes of the tensor lies along the Fe-As bond axis, then we can constrain $A_{ab}/A_{aa} = 0.37$ or $-0.94$, thus it is reasonable to assume $A_{ab}<A_{aa}$ .

We are now in a position to analyze the experimental results displayed
in Fig. \ref{fig:T1anisotropy}. The presence of tensile strain $\varepsilon$
along the $a$ axis effectively induces a conjugate field that couples
to the nematic order parameters, i.e. $\varphi_{i}\propto\varepsilon$.
As a result, the nematic phase extends to high temperatures, and $T_{S}$
signals a crossover rather than an actual phase transition. Furthermore,
because in our experiment tensile strain is applied along the $a$
axis, the $\mathbf{Q}_{1}$ ordering vector is selected by the external
strain, with spins pointing along the $a$ axis (see Fig. \ref{fig:xtalphotos}
and also Ref. \onlinecite{StrainedPnictidesNS2014science}). As a result,
$\varphi_{1}>0$, although $\varphi_{2}$ and $\varphi_{3}$ could
in principle have different signs.

At temperatures much larger than the magnetic transition temperature
$T_{N}$, the effects of the spin-orbit coupling are presumably small.
Therefore, in this regime, the magnetic spectrum should display an
isotropic behavior, with $\xi_{i}\approx\xi$. In this regime the
last term in Eq. (\ref{eqn:T1TinvANIS}) dominates, and the sign of
$(T_{1}T)_{{\rm anis}}^{-1}$ is the opposite as the sign of $\varphi_{3}$.
According to the data plotted in Fig. \ref{fig:T1anisotropy}(b),
within the experimental error bars, $(T_{1}T)_{{\rm anis}}^{-1}<0$
at high temperatures, suggesting that $\varphi_{3}>0$.

As the magnetic transition is approached, the effects of the spin-orbit
coupling presumably become more important. In particular, because
in the magnetically ordered state the magnetic moments point parallel
to the ordering vector $\mathbf{Q}_{i}$, $\xi_{1}$ must be the only
correlation length that diverges at the magnetic transition, i.e.
$\xi_{1}\gg\xi_{2},\xi_{3}$ at $T\gtrsim T_{N}$. Consequently, the
first term in Eq. (\ref{eqn:T1TinvANIS}) should dominate in this
regime. Because $\varphi_{1}>0$ and $A_{ab}<A_{aa}$, we expect that
$\left(T_{1}T\right)_{\mathrm{anis}}^{-1}<0$ near the transition.
This expectation, however does not agree with the observed behavior
seen in Fig. \ref{fig:T1anisotropy}(b).

We can also analyze the isotropic response, $\Delta\left(T_{1}T\right)_{\mathrm{iso}}^{-1}$.
According to Eq. (\ref{eqn:T1TinvISO}), $\Delta\left(T_{1}T\right)_{\mathrm{iso}}^{-1}$
is always positive. Indeed, neutron scattering experiments in both
twinned\cite{Zhang2015} and detwinned\cite{BaCo122TNstrainedPRB2014} samples find
enhanced magnetic fluctuations in the nematic phase. However, our
data presented in Fig. \ref{fig:T1anisotropy}(b) shows that $\Delta\left(T_{1}T\right)_{\mathrm{iso}}^{-1}$
is positive only at high temperatures -- roughly within the same regime
in which $\left(T_{1}T\right)_{\mathrm{anis}}^{-1}<0$ -- and becomes
negative as $T_{N}$ is approached, in contrast with the prediction
of Eq. (\ref{eqn:T1TinvISO}).

There are several possible reasons for this discrepancy between the
theoretical calculation and the observed data in the temperature regime
near $T_{N}$, including: (i) unequal strain between the two different
field orientations; (ii) crystal misalignments; and (iii) higher order
corrections due to non-infinitesimal strain. Note that this device
nominally applies a constant stress, rather than constant strain,
and differential thermal contraction between the mechanical clamp,
the silver wires, and the sample likely leads to a temperature-dependent
induced strain. Because the nematic order parameters $\varphi_{i}$
should be proportional to the strain, these quantities may not be
the same for the two different field directions in the measured values.
For example, if the wires used to suspend the sample exhibit a temperature-hysteretic
effect due to thermal contractions that exceed the elastic regime,
then the strain applied for the two different directions will be different.

Scenario (ii), crystal misalignment, could arise from a small component
of $\mathbf{H}_{0}$ along the $c$ direction that is different between
different crystal orientations, which would contribute an asymmetry
that would not cancel out. As the crystal is suspended in free space
by the wires, it is possible that differences in thermal expansions
could lead to torques that could twist the crystal, giving rise to
a difference between the crystal orientation between strained and
unstrained conditions. Detailed studies of the NMR spectra (not shown) in the
ordered state of undoped BaFe$_{2}$As$_{2}$ under strain in this
device indicate that misalignments of 1-2$^{\circ}$ are possible.

The third scenario, namely higher-order strain-induced changes to
the \slrrtext, could be present depending on the sensitivity of the
nematic order parameters, $\varphi_{i}$, to strain. Nominally, the
applied strains are small, and are just enough to detwin the crystal.
We observe little or no shift in the peak of $(T_{1}T)_{\alpha}^{-1}$
at $T_{N}$ in Fig. \ref{fig:rawT1data}, suggesting that the main
effect of strain is to detwin the crystal. However, for sufficiently
high strain levels, $T_{N}$ is known to increase,\cite{BaCo122TNstrainedPRB2014}
and therefore the temperature dependence of the correlation lengths,
$\xi_{i}$ will be altered.

In this regard, we note that the theoretical analysis presented here
considers the linear response of $(T_{1}T)_{{\rm anis}}^{-1}$ to
strain. From generic symmetry considerations, in the linear-response
regime, one expects that $\Delta(T_{1}T)_{a}^{-1}$ and $\Delta(T_{1}T)_{b}^{-1}$
display opposite behaviors. From Fig. \ref{fig:T1anisotropy}, this
does seem to be the case at higher temperatures, where in fact the
theoretical predictions for $(T_{1}T)_{{\rm anis}}^{-1}$ and $\Delta(T_{1}T)_{{\rm iso}}^{-1}$
are in qualitative agreement with the data. As $T_{N}$ is approached,
however, both $\Delta(T_{1}T)_{a}^{-1}$ and $\Delta(T_{1}T)_{b}^{-1}$
display the same behavior, indicating the onset of non-linear effects
beyond the analysis presented here. To mitigate these issues, it would
be interesting to control precisely the applied strain using a piezo
device, as it was done in Ref. \onlinecite{FisherScienceNematic2012} for
resistivity measurements.

\section{Conclusions}

In summary, we have measured the NMR spin-lattice relaxation rate
in Ba(Fe$_{1-x}$Co$_{x}$)$_{2}$As$_{2}$ with $x=0.048$ under
uniaxial tensile stress as a function of temperature, and found significant
changes to the relaxation rate that persist above $T_{s}$ in a detwinned
crystal. The strain field breaks $C_{4}$ symmetry, and the anisotropic
magnetic fluctuations probed by \slrr reflect the impact of nematicity
on the fluctuation spectrum. Surprisingly, the glassy behavior manifested
by the broad distribution of relaxation times is unaffected under
strain. This observation suggests that the local strains, introduced
either by the Co dopants or by lattice defects, exceed the applied
strain. Consequently the glassy behavior is not associated with large
nematic domains.

We also compute the spin-lattice relaxation rate using a model for
the anisotropic dynamical spin susceptibility. By introducing nematic
order parameters that reflect the changes to the spin-spin correlation
lengths along the three crystal axes, we estimate the leading contributions
to the anisotropy of the spin-lattice relaxation rate in the presence
of strain. Theoretically, we find that the strain-induced changes
to $(T_{1}T)_{a,b}^{-1}$ should have opposite signs. On the other
hand, experimentally we find that this is the case only at high temperatures,
since both quantities are suppressed as $T_{N}$ is approached. This
discrepancy most likely arises due to crystal misalignments between
the strained- and unstrained states, and/or differences in induced
strains between the two different directions. Future measurements
with more precise control over the orientation and amplitude of the
strain will provide detailed information about the relative sizes
of the nematic order parameters, $\varphi_{i}$, under strain. Nevertheless,
our experiments show that the combination of NMR and strain is a unique
tool to probe not only the effect of nematic order on the unpolarized
magnetic spectrum, but most importantly on the polarized spin spectrum,
revealing the interplay between nematicity and spin-orbit coupling.

\begin{acknowledgements} We thank S. Kivelson, E. Carlson, P. Pagliuso
and R. Urbano for enlightening discussions, and P. Klavins, J. Crocker,
K. Shirer, and M. Lawson for assistance in the laboratory. Work at
UC Davis was supported by the NSF under Grant No.\ DMR-1506961. RMF
is supported by the U.S. Department of Energy, Office of Science,
Basic Energy Sciences, under award number DE-SC0012336. Work done
at Ames Lab (SLB, PCC, MT, RP, EIT) was supported by the U.S. Department
of Energy, Office of Basic Energy Science, Division of Materials Sciences
and Engineering. Ames Laboratory is operated for the U.S. Department
of Energy by Iowa State University under Contract No. DE-AC02-07CH11358. This work was completed at the Aspen Center for Physics, which is supported by the National Science Foundation grant PHY-1066293.
\end{acknowledgements}

\bibliography{Ba122_Co-doped_strain}

\begin{widetext}
\appendix

\section{Details of the calculation of the anisotropic spin-lattice relaxation
rate}
\label{sec:app}

Eq. (\ref{def}) of the main text contains the hyperfine tensor:

\[
\bar{A}_{\mathbf{q}}=4\left(\begin{array}{ccc}
A_{aa}\cos\left(\frac{q_{x}}{2}\right)\cos\left(\frac{q_{y}}{2}\right) & -A_{ab}\sin\left(\frac{q_{x}}{2}\right)\sin\left(\frac{q_{y}}{2}\right) & iA_{ac}\sin\left(\frac{q_{x}}{2}\right)\cos\left(\frac{q_{y}}{2}\right)\\
-A_{ba}\sin\left(\frac{q_{x}}{2}\right)\sin\left(\frac{q_{y}}{2}\right) & A_{bb}\cos\left(\frac{q_{x}}{2}\right)\cos\left(\frac{q_{y}}{2}\right) & iA_{bc}\cos\left(\frac{q_{x}}{2}\right)\sin\left(\frac{q_{y}}{2}\right)\\
iA_{ca}\sin\left(\frac{q_{x}}{2}\right)\cos\left(\frac{q_{y}}{2}\right) & iA_{cb}\cos\left(\frac{q_{x}}{2}\right)\sin\left(\frac{q_{y}}{2}\right) & A_{cc}\cos\left(\frac{q_{x}}{2}\right)\cos\left(\frac{q_{y}}{2}\right)
\end{array}\right)
\]
and the rotation matrix:

\[
\bar{R}=\left(\begin{array}{ccc}
\sin^{2}\phi+\cos\theta\,\cos^{2}\phi & -\sin2\phi\,\sin^{2}\frac{\theta}{2} & \cos\phi\,\sin\theta\\
-\sin2\phi\,\sin^{2}\frac{\theta}{2} & \cos^{2}\phi+\cos\theta\,\sin^{2}\phi & \sin\phi\,\sin\theta\\
-\cos\phi\,\sin\theta & -\sin\phi\,\sin\theta & \cos\theta
\end{array}\right)
\]

Using Eq. (\ref{def}), we can also calculate $1/T_{1}T$ for a field
applied parallel to $c$:

\begin{eqnarray}
\left(\frac{1}{T_{1}T}\right)_{c} & = & \frac{g^{2}}{2}\sum_{\mathbf{q}}\left\{ \left[\sin^{2}\left(\frac{q_{x}}{2}\right)\sin^{2}\left(\frac{q_{y}}{2}\right)A_{ab}^{2}+\cos^{2}\left(\frac{q_{x}}{2}\right)\cos^{2}\left(\frac{q_{y}}{2}\right)A_{aa}^{2}\right]\left[\tilde{\chi}_{aa}\left(\mathbf{q}\right)+\tilde{\chi}_{bb}\left(\mathbf{q}\right)\right]\right.\nonumber \\
 &  & \left.+\frac{1}{2}\left[1-\cos\left(q_{x}\right)\cos\left(q_{y}\right)\right]A_{ac}^{2}\tilde{\chi}_{cc}\left(\mathbf{q}\right)\right\} \label{T1T_c}
\end{eqnarray}
as well as the isotropic response $\left(T_{1}T\right)_{\mathrm{iso}}^{-1}=\left(T_{1}T\right)_{a}^{-1}+\left(T_{1}T\right)_{b}^{-1}$:

\begin{eqnarray}
\left(\frac{1}{T_{1}T}\right)_{a}+\left(\frac{1}{T_{1}T}\right)_{b} & = & \frac{g^{2}}{2}\sum_{\mathbf{q}}\left\{ \left[\sin^{2}\left(\frac{q_{x}}{2}\right)\sin^{2}\left(\frac{q_{y}}{2}\right)A_{ab}^{2}+\cos^{2}\left(\frac{q_{x}}{2}\right)\cos^{2}\left(\frac{q_{y}}{2}\right)A_{aa}^{2}+\right.\right.\nonumber \\
 &  & \left.+\frac{1}{2}\left(1-\cos q_{x}\cos q_{y}\right)A_{ca}^{2}\right]\left(\tilde{\chi}_{aa}\left(\mathbf{q}\right)+\tilde{\chi}_{bb}\left(\mathbf{q}\right)\right)-\left[\frac{1}{2}\left(\cos q_{x}-\cos q_{y}\right)A_{ca}^{2}\right]\left(\tilde{\chi}_{aa}\left(\mathbf{q}\right)-\tilde{\chi}_{bb}\left(\mathbf{q}\right)\right)\nonumber \\
 &  & \left.+\left[\frac{1}{2}\left(1-\cos q_{x}\cos q_{y}\right)A_{ac}^{2}+2\cos^{2}\left(\frac{q_{x}}{2}\right)\cos^{2}\left(\frac{q_{y}}{2}\right)A_{cc}^{2}\right]\tilde{\chi}_{cc}\left(\mathbf{q}\right)\right\} \label{T1T_ab}
\end{eqnarray}

Focusing on the behavior at the magnetic ordering vectors $\mathbf{Q}_{1}=\left(\pi,0\right)$
and $\mathbf{Q}_{2}=\left(0,\pi\right)$, we note that $\left(T_{1}T\right)_{c}^{-1}$
is dominated by the out-of-plane fluctuations, $\chi_{cc}\left(\mathbf{Q}_{i}\right)$,
whereas $\left(T_{1}T\right)_{\mathrm{iso}}^{-1}$ contains also contributions
from $\chi_{aa}\left(\mathbf{Q}_{1}\right)$ and $\chi_{bb}\left(\mathbf{Q}_{2}\right)$.\end{widetext}

\end{document}